# Superconducting $Fe_{1+\delta}Se_{1-x}Te_x$ thin films: Growth, characterization and properties

T. Geetha Kumary,* Dipak Kumar Baisnab, J. Janaki, Awadhesh Mani, R. M. Sarguna, P. K. Ajikumar, A. K. Tyagi and A. Bharathi

Materials Science Group, Indira Gandhi Centre for Atomic Research, Kalpakkam, Tamilnadu – 603 102, India

## Abstract

Thin films of $Fe_{1+\delta}Se_{1-x}Te_x$ ($\delta \sim 0.18$ & $x \sim 0.5$) have been successfully grown on (100) oriented single crystalline $SrTiO_3$ and $LaAlO_3$ substrates by pulsed laser deposition. The crystal structure was characterized by x-ray diffraction, and the superconducting properties by electrical resistivity measurements. X-ray diffraction analysis establishes the epitaxial growth of the films with c-axis orientation. Atomic force microscopy showed a smooth surface morphology for the films grown on $SrTiO_3$ substrates. All the films are observed to be superconducting with a $T_c$ of ~ 8-14 K depending on the deposition conditions. The deposition parameters were optimized to obtain good quality films with $T_c$ comparable to that of the target material.

---

* Corresponding author; email: geetha@igcar.gov.in

**INTRODUCTION**

The iron based chalcogenide superconductors [1, 2] as well as the iron based pnictide superconductors [3, 4] discovered recently have ushered a new era in the field of superconductivity. Both the iron pnictides and chalcogenides have P4/nmm structure with superconducting FeAs (or FeSe) layers having the anti-fluorite structure in which Fe is tetrahedraly co-ordinated to As (or Se) with the only difference that in the arsenides there are charge reservoir layers as well. The undoped parent phase of the Fe-As based superconductors is magnetically ordered with a spin density wave (SDW) state, which gets suppressed on electron or hole doping with the emergence of superconductivity [5]. On the other hand, FeSe superconducts at a $T_c \sim 8$ K, in the undoped form and is reported to have similar electronic properties as the electron doped Fe-As superconductors [6].

Application of external pressure as well as partial substitution of Se by Te or S leads to an increase [7-10] in $T_c$ in FeSe. While the increase in $T_c$ with S doping can be reconciled with application of chemical pressure, the increase in $T_c$ due to Te is attributed [7] to the higher stability of the SDW state in FeTe compared to FeSe [11]. Another effect in favour of Te substitution is that the tetragonal phase, crucial for the occurrence of superconductivity in this system, is stabilized by Te substitution. For example, $Fe_{1+y}Te$ is stabilized over a composition range [12] y ~ 0.06-0.17, whereas the tetragonal structure in $Fe_{1+x}Se$ is stabilized in a narrower composition window of x=0.01- 0.025, and is extremely sensitive to the synthesis conditions [13]. Solid solution of tetragonal $Fe_{1+\delta}Se_{1-x}Te_x$ forms for $0 \leq x \leq 1$ with a maximum $T_c$ (~14 K) and superconducting volume fraction for x ~ 0.5 [9, 10, 14]. All theses studies have been carried out on essentially phase pure polycrystalline materials, although some single crystal studies [14-16] also exists. Detailed studies to unravel the intrinsic nature of superconductivity in materials require single crystals. Study of thin films also provides another viable method to investigate the intrinsic anisotropic superconducting properties. A few studies of superconducting thin films (Co doped $SrFe_2As_2$) have been reported [17, 18] in the case of iron arsenide superconductors. Also, undoped $SrFe_2As_2$ thin film has been shown to become superconducting by water intercalation [19]. However, growth and study of superconducting thin films of iron chalcogenides have not yet been reported in literature and would be of much interest from the point of view of basic research as well as practical applications. In the present paper, we report on the first attempt of the growth and characterization of thin films of the superconducting iron chalcogenide pseudo-binary system $Fe_{1+\delta}Se_{1-x}Te_x$ ($\delta \sim 0.18$ and $x \sim 0.5$). Our results indicate that epitaxial superconducting thin films can be successfully grown by Pulsed Laser Deposition.

**EXPERIMENTAL DETAILS**

Polycrystalline pellets of target material of nominal composition $Fe_{1+\delta}Se_{1-x}Te_x$ with $\delta=0.18$ & $x = 0.5$ used for thin film preparation, were synthesized by solid state reaction of the high purity constituent elements as described in Janaki *et. al.* [20]. Thin films were prepared by using a pulsed laser deposition system wherein a KrF excimer laser of 248 nm wavelength is used, with a laser fluence of ~1.5 J/cm$^2$. The films were grown on 1cm$^2$, single crystalline substrates of $LaAlO_3$ (LAO) and $SrTiO_3$ (STO) with (100) orientation. These substrates were chosen because they have comparable lattice parameters with the *a* lattice parameter of $Fe_{1.18}Se_{0.5}Te_{0.5}$. Thin film deposition was carried out under vacuum (~ 10$^{-3}$ mbar) to avoid oxide formation. Substrate temperature was maintained at 400ºC during deposition and the films were annealed at this temperature for 30 minutes, in-situ, after deposition. Higher substrate temperature was avoided as the non-superconducting hexagonal phase was seen to form (for T > 450ºC) in preference to the superconducting tetragonal phase in the synthesis of polycrystals of FeSe [20]. The deposition parameters such as the laser pulse frequency and target to substrate distance were varied to optimize the quality of the films. The films designated as FSTLAO1 and FSTSTO3 were deposited under identical conditions on LAO and STO substrates respectively, by keeping a distance of 3.5 cm between the target and substrate and by keeping a laser frequency of 10

Hz (see Table 1). On the other hand a laser frequency of 5 Hz and target to substrate distance of 4 cm is employed during the preparation of the films, FSTLAO2 and FSTSTO4 (on LAO and STO substrates, respectively). The thicknesses of the thin films were measured using a Dektak surface profiler. The crystal structure was characterized by x-ray diffraction (XRD) measurements in a STOE diffractometer using Cu $K_\alpha$ radiation. The surface morphology of the films was studied using Atomic Force Microscopy (AFM). Electrical resistivity measurements were carried out using the four probe technique in a dipstick-cryostat for the bulk and thin film samples.

**RESULTS AND DISCUSSION**

Visual inspection of the films indicated the presence of mirror like shiny surfaces for all the films. The thicknesses of FSTLAO1 and FSTSTO3 were measured to be ~250 nm whereas FSTLAO2 and FSTSTO4 were found to have slightly lower thicknesses of ~120 nm and 100 nm respectively (see Table 1), due to the larger target to substrate distances used in preparing the later films. Figures 1(a) – (d) represent the XRD patterns of all the films in a logarithmic scale. Occurrence of only the (*00l*) lines of the tetragonal FeSe structure with the [100] reflections of the matrix indicate that the thin films have *c*-axis orientation. A close examination of the XRD pattern of the film FSTLAO1 reveals the appearance of double peaks for all the (*00l*) reflections, indicating the presence of two phases possibly with different Se/Te ratios. In addition, a few low intensity impurity peaks due to the presence of $SeO_2$, FeSe and γ-Fe are also seen in the XRD pattern (see. Fig.1(a)). The diffraction pattern of the film grown under the same conditions on the STO substrate (FSTSTO3) is shown in Fig. 1(b). The XRD pattern indicates that the quality of the film is improved compared to that grown on LAO as the (*00l*) peaks are not split, indicating a more uniform Se/Te composition in the film. This film also, however contains a small quantity of $SeO_2$ and FeO impurities as indicated in XRD pattern. The films FSTLAO2 and FSTSTO4 grown at a lower deposition rate are observed to have better XRD patterns with lesser impurity lines. FSTLAO2 has no impurity peaks while FSTSTO4 shows a small intensity corresponding to FeO impurity. The *c* lattice parameters estimated from the (001) reflection for all the films and are given in Table 1 along with that of the bulk polycrystalline sample. The grain sizes calculated from XRD patterns of the samples, using the Scherrer formula after subtracting the instrumental broadening from the observed line broadening, are displayed in table 1 along with all the other parameters measured on the thin films.

Atomic Force Micrographs were taken in order to study the surface morphology of the thin films. The surface roughnesses are observed to be larger for the thin films grown on LAO substrates. A maximam roughness of ~ 50 nm is measured for FSTLAO1 while FSTLAO2 has a smoother surface with roughness of ~22 nm. On the other hand the films grown on STO showed a much smoother surface with maximum roughness of ~ 6 nm. The representative AFM pictures of two thin films, FSTLAO1 and FSTSTO3 are shown in Figs. 2(a) and (b), respectively. The two dimensional surface showing the individual grains for one of the thin films, FSTSTO4 is shown in Fig. 3. The grain size is of the order of 100 nm in this film, which is consistent with the results obtained from x-ray analysis.

The temperature dependent resistivity, $\rho(T)$ in the 4.2 K – 300 K temperature range for all samples are shown in Fig. 4 with the region around the transition magnified in the inset. As can be seen from the inset of Fig. 4 both the thin films FSTLAO1 and FSTSTO3 exhibit zero resistivity with a $T_c$ (defined as the temperature at which the resistivity deviates from the normal state behavior) of ~ 8 K. But the normal state resistivity of the FSTLAO1 shows a large negative coefficient of resistivity in the normal state, characteristic of the samples with Te excess [14]. The normal state resistivity in the FSTSTO3 sample shows metallic resistivity in the 25 K – 300 K temperature range (see Fig.4) but shows a strong upturn just above the superconducting transition, which is again displayed by samples with Te excess [14] or with poor intergrain connectivity. The room temperature resistivity is ~1.6 mΩ-cm in both

FSTLAO1 and FSTSTO3. The films grown with a lower deposition rate (FSTLAO2 and FSTSTO4) exhibit higher $T_c$ values (see Table 1) although the transition widths are larger than that obtained at faster deposition rates in FSTLAO1 and FSTSTO3. FSTLAO2 grown on LAO substrate has a $T_c$ of ~11.5 K whereas FSTSTO4 grown on STO shows even higher $T_c$ of ~ 14 K. It is remarkable that the normal state resistivity is more or less flat with respect to temperature in both FSTLAO2 and FSTSTO4. The resistivity at room temperature shows a significant decrease in these samples (see table 1) and is very close to values reported in the single crystals [14]. The low value of room temperature resistivity and the nearly temperature independent resistivity observed in the normal state of FSTLAO2 and FSTSTO4 thin films indicate that they are of good quality. For comparison, the resistivity behavior in the 4 K – 300 K temperature range of the polycrystalline target material $Fe_{1.18}Se_{0.5}Te_{0.5}$ is shown in Fig. 5. The region near the superconducting transition is shown in the inset in an expanded scale. While the room temperature resistivity (see table 1) in the polycrystal is small like that obtained in the thin films FSTLAO2 and FSTSTO4, the $\rho(T)$ shows a change in slope at ~125 K. This can either be due to a small inhomogeneity in the Se/Te ratio or due to the presence of small amount of the $Fe_3O_4$ impurity phase present in the polycrystalline sample [20]. All parameters measured from the resistivity data on the various thin films grown in this study along with that of the polycrystalline sample are summarized in Table 1. A careful inspection of Table 1 indicates that higher $T_c$ correlates with the lower room temperature resistivity. Smoother films are grown on STO compared to LAO substrates. A slower deposition rate favors the formation of films with reduced disorder and higher $T_c$. From Fig. 4 it is apparent that resistivity behavior in the thin films FSTLAO2 and FSTSTO4 are closer to the behavior seen in single crystalline $FeSe_{0.5}Te_{0.5}$, as compared to that observed in the target (see fig.5) implying that by choosing appropriate deposition conditions, a thin film with better transport characteristics than that of the target can be produced.

As is apparent from table 1, the *c* lattice parameter is found to be around ~ 5.9 Å for all the films and is slightly lower than that observed for the target material (5.9710 Å). It doesn't show any correlation to the observed $T_c$ variations. Variation in $T_c$ can arise due to a difference in Se/Te ratio [9] or due to a strain in the lattice. Since higher $T_c$ is observed for films with lower thicknesses (~100 nm) the effect of strain should be negligible in these films. The large $T_c$ value of 14 K in the FSTSTO4 sample seems to point to a Se/Te ratio [9] of ~ 1:1. Compositional analyses carried out using Energy Dispersive Spectroscopy in different regions of the bulk polycrystalline sample resulted in a composition of around $Fe_{1.1}Se_{0.55}Te_{0.45}$. However, the composition of thin films could not be obtained due to the poorer sensitivity of the equipment for thin films.

## SUMMARY

Superconducting thin films of $Fe_{1+\delta}Se_{1-x}Te_x$ ($\delta \sim 0.18$ & $x \sim 0.5$) are grown on $LaAlO_3$ and $SrTiO_3$ substrates by pulsed laser deposition. XRD analysis shows that the films are grown epitaxially with c-axis orientation. Films grown on $SrTiO_3$ have smoother surface morphology compared to those grown on $LaAlO_3$ as observed from the atomic force micrographs. Thin films of $Fe_{1+\delta}Se_{1-x}Te_x$ with a $T_c$ of ~14 K, that is comparable to that of the bulk, could be produced by optimizing the laser deposition conditions.

## ACKNOWLEDGMENT

The authors acknowledge Dr. S. Kamruddin, Materials Science Group, Indira Gandhi Center for atomic Research, Kalpakkam for the help in AFM measurements.

Table 1: Thicknesses, roughness, grain size, $c$ lattice parameters, $T_c$ and room temperature resistivity ($\rho(RT)$) of the samples studied for the various conditions of deposition for a target temperature of $400^0C$

| Sample | Thickness (nm) | Roughness (nm) | Grain size | $c$-parameter (Å) | $T_c$ (K) | $\rho(RT)$ (mΩ cm) | Target to substrate distance (cm) | Laser Pulse frequency (Hz) |
|---|---|---|---|---|---|---|---|---|
| Bulk | – | – | – | 5.9710 | 13.5 | 1.18 | – | – |
| FSTLAO1 | 250 | 50 | 50 | 5.9062 | 8 | 1.582 | 3.5 | 10 |
| FSTSTO3 | 250 | 6 | 80 | 5.8886 | 8 | 1.682 | 3.5 | 10 |
| FSTLAO2 | 100 | 25 | 100 | 5.9027 | 11.5 | 1.11 | 4.0 | 5 |
| FSTSTO4 | 120 | 6 | 100 | 5.8999 | 14 | 0.912 | 4.0 | 5 |

**Figure Captions**

Fig. 1 X-ray diffraction patterns of the thinfilms grown on LaAlO$_3$ substrates (a) FSTLAO1 (c) FSTLAO2 and on SrTiO$_3$ substrates (b) FSTSTO3 and (d) FSTSTO4 (carried out under different deposition conditions explained in the text). The low intensity impurities present are indicated.

Fig.2 Atomic force micrographs of (a) FSTLAO1 and (b) FSTSTO3. Film grown on SrTiO$_3$ has a smoother surface than that grown on LaAlO$_3$ substrate.

Fig. 3 Two dimensional surface morphologhy obtained from AFM of FSTSTO4 showing grains of sub-micron size.

Fig.4 Resistivity versus temperature for thin film samples in the 4.2 K - 300K range. Inset shows the resistivity close to the superconducting transitions.

Fig. 5 Resisitivity as a function of temperature for the bulk polycrystalline target. Region near the superconducting transition is shown in the inset.

Fig. 1

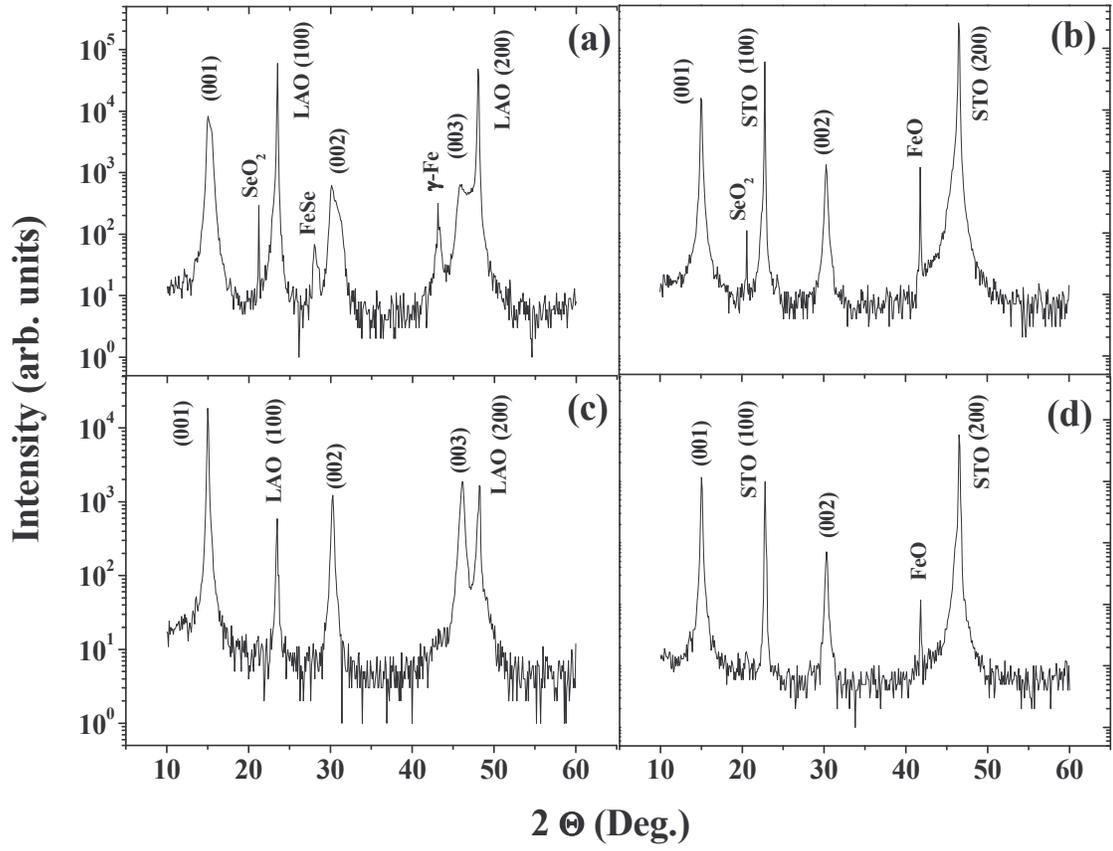

**Fig.2**

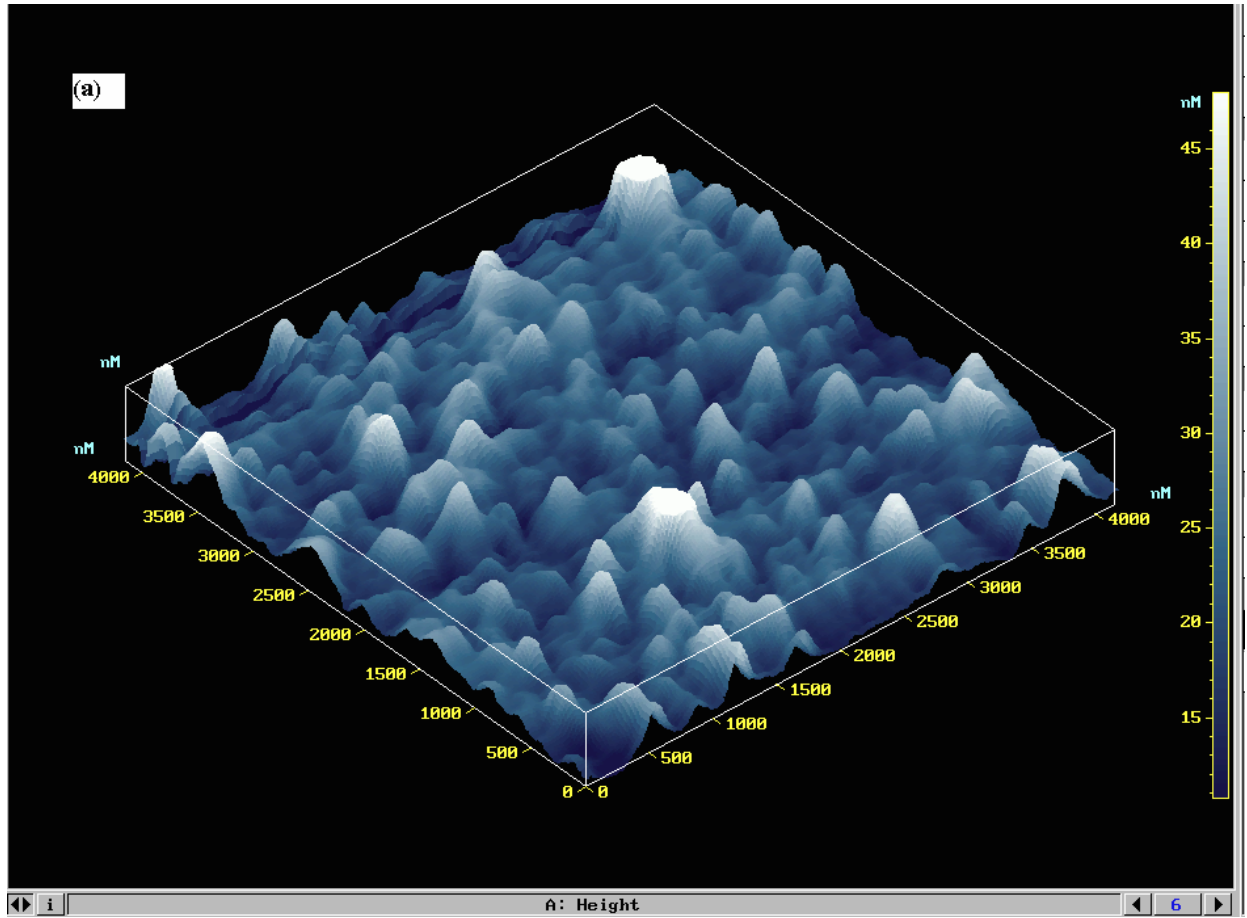

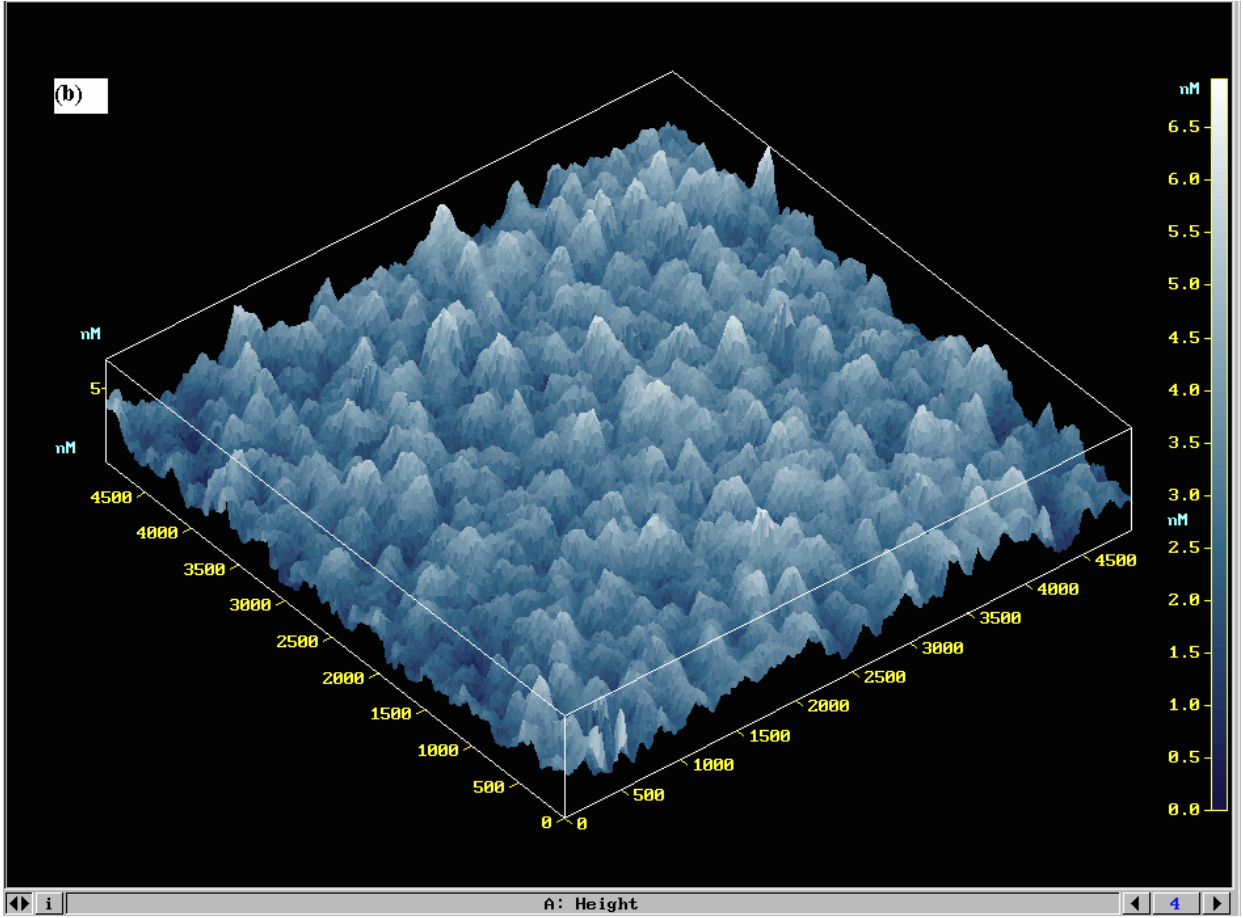

**Fig. 3**

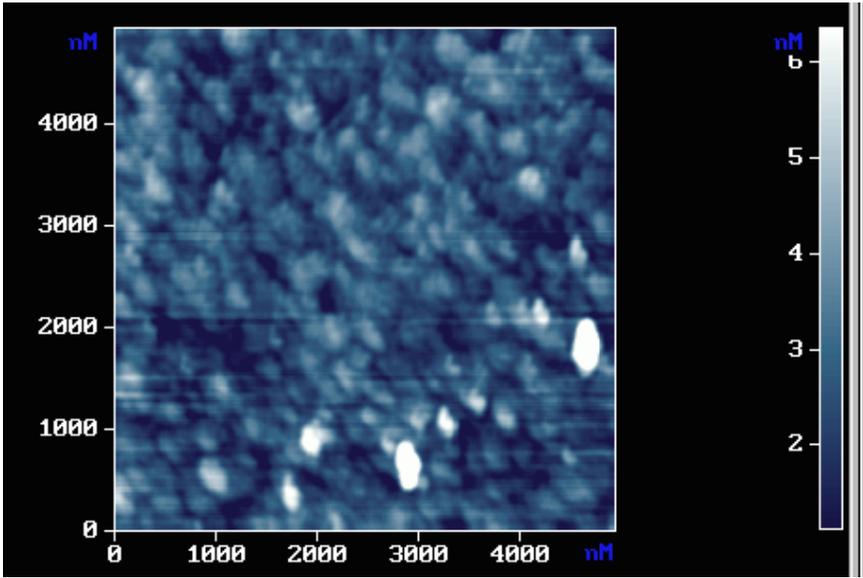

Fig. 4

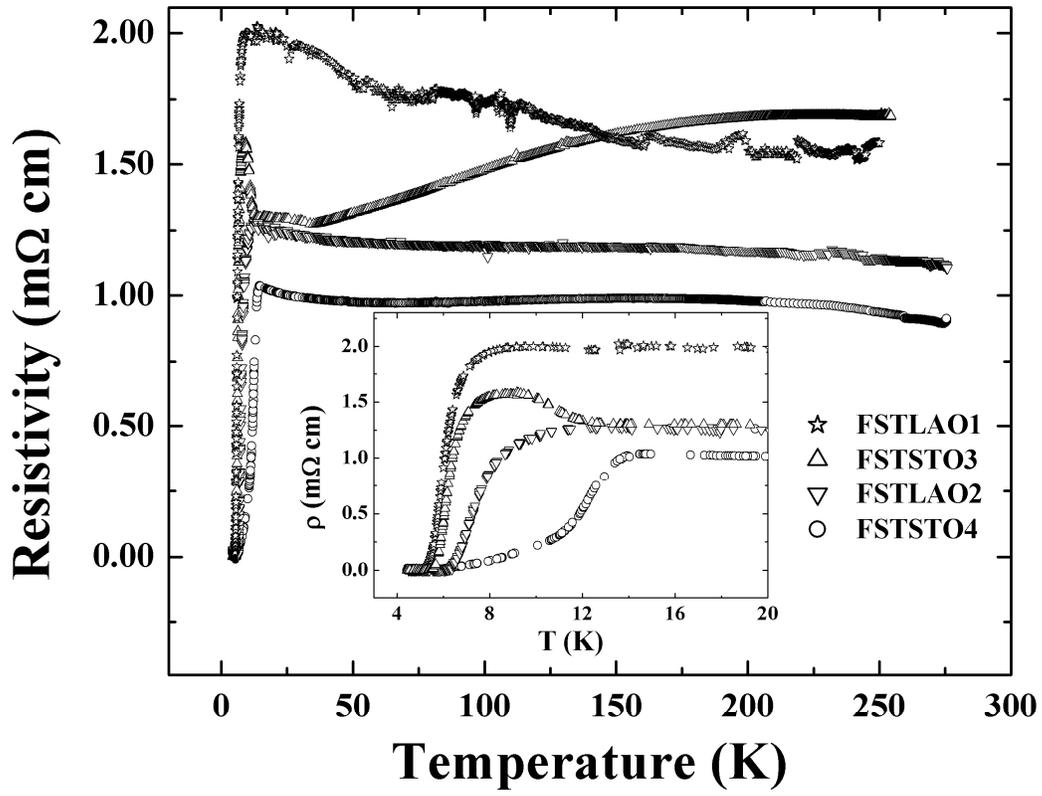

Fig. 5

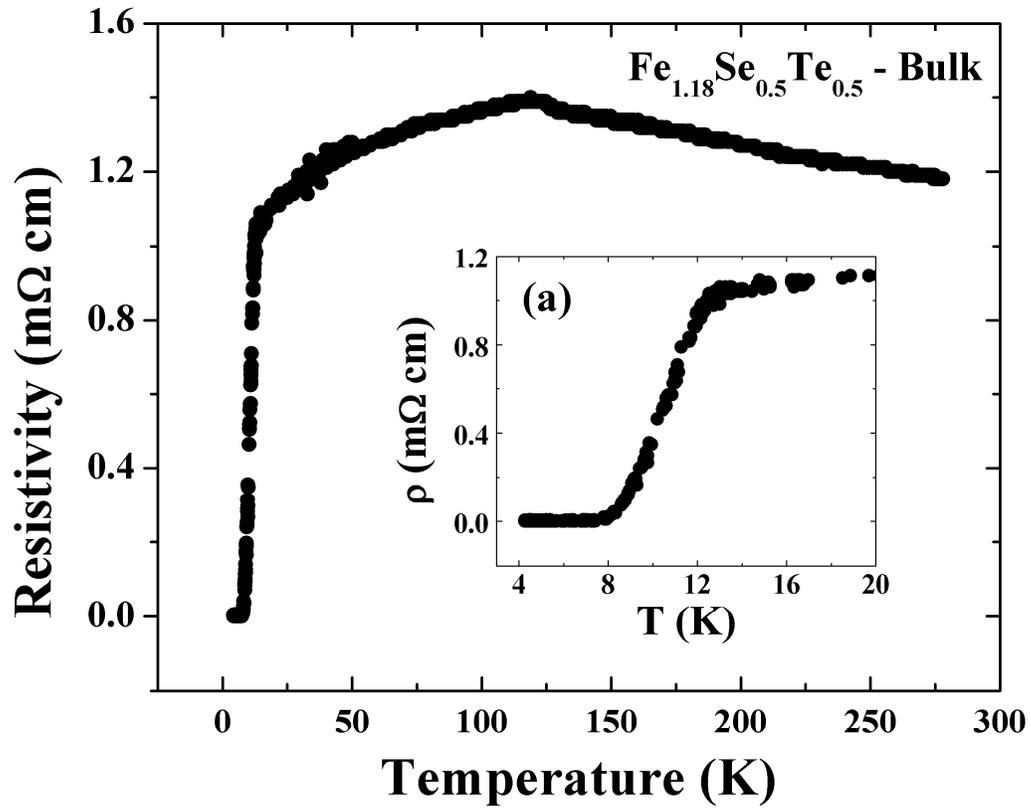